\documentclass[aps,showkeys,superscriptaddress]{revtex4}

\begin{document}
\title{Effective action for strongly correlated electron systems}
\author{A. Ferraz}
\affiliation{International Institute of Physics - UFRN,
Department of Experimental and Theoretical Physics - UFRN, Natal, Brazil}
\author{E.A. Kochetov}
\affiliation{International Institute of Physics - UFRN, Natal, Brazil;\\
Laboratory of Theoretical Physics, Joint
Institute for Nuclear Research, 141980 Dubna, Russia}

\begin{abstract}

The $su(2|1)$ coherent-state path-integral representation of the partition function
of the $t-J$ model of strongly correlated electrons
is derived at finite doping.
The emergent effective action is compared to the one
proposed earlier on phenomenological grounds by Shankar to describe holes in an antiferromagnet
(Nucl.Phys. B330 (1990) 433).
The $t-J$ model effective action is found
to have an important "extra" factor with no analogue
in Shankar's action.
It represents the local constraint of no double electron occupancy and reflects the
rearrangement of the underlying  phase-space manifold due to the presence of strong electron correlation.
This important ingredient is shown to be essential to describe
the physics of strongly correlated electron systems.

\end{abstract}

\maketitle
\section{Introduction}

In this work, we discuss a path-integral representation of the partition function
for strongly correlated electron systems. In particular, we are interested in the
effective low-energy action to describe a lightly doped antiferromagnet (AF).
Strong electron correlations essentially determine the low-energy physics of high-T$_c$
superconductors \cite{anderson},
heavy-fermion systems \cite{coleman}, itinerant ferromagnets \cite{nagaoka}, as well as some optical lattices \cite{trap}.
Accordingly, many various approaches have been proposed to study those corresponding low-energy effective actions.
We will comment on some of them at the end of the paper.

Our work has actually been motivated by Shankar's contribution \cite{shankar}.
Namely, two decades ago Shankar put forward a conjecture assuming that the low-energy
action to describe a doped quantum AF involves spinless  fermions locally
coupled to a compact $U(1)$ gauge field. This gauge field is  driven by
the AF fluctuations controlled by a nonlinear sigma-model.
This approach was further discussed in \cite{dorey}.
A natural question then arises as to whether that action can be derived directly from the $t-J$ model
for strongly correlated electrons at finite doping. After all, this microscopical model is widely believed to
capture the essential physics of a lightly doped AF.

In the present paper  we show that Shankar's effective action
can indeed be derived from
the $su(2|1)$ path-integral representation of the partition function of the $t-J$ model.
However, the resulting path-integral measures differ by an important factor.
This distinction is a manifestation of the strong coupling
nature of the problem due to the no double occupancy
(NDO) constraint. It may seem that this constraint is already fully accounted in Shankar's theory
by the requirement that the fermions are spinless, since no double fermion occupation is possible
in this case.
However, the NDO constraint modifies the original on-site Hilbert space as well as
the on-site  path-integral  phase space, which is not explicitly taken into consideration in \cite{shankar}.
Shankar proposed instead an effective action right in terms of conventional fermion and spin
fields to describe a doped AF. This of course implies a standard measure in the path integral.
Although, his action may presumably describe some unconstrained spin-fermion models, it is not
appropriate for strongly correlated electrons.

Our aim is to demonstrate in what way the NDO constraint
modifies the theory discussed in \cite{shankar}.
To start with, we illustrate our point by considering a path-integral representation
of the partition function for a simple single-site Hamiltonian.
After that, we address  the $t-J$ model close to half filling which is precisely
the physically most relevant situation
for strongly correlated electrons.
This model is believed to capture the low-energy physics of lightly
doped quantum AF.

Quite plainly,
an electron system is said to be strongly correlated  if the leading energy scale in the problem
is the on-site Coulomb repulsion energy $U$. In this case the low-energy sector of the underlying
on-site Hilbert space should be
modified to exclude doubly occupied states. Such a modification results in an entirely new
physics to account for the relevant low-energy excitations.
Formally, strong correlations are encoded into the projected electron (Hubbard) operators. They act directly
in the restricted Hilbert space as opposed to the conventional electron operators which
describe the unconstrained system.
In contrast with the conventional fermion operators which generate  the standard  fermionic algebra, the new  operators
obey more complicated commutation/anticommutation relations and are closed into a superalgebra $su(2|1)$.
It is therefore natural
to seek a path-integral
formalism that takes the structure of that superalgebra fully into consideration. In  analogy with the conventional
$su(2)$ spin path integral~\cite{klauder}, this can be carried out by employing an appropriate coherent-state
basis associated with the $su(2|1)$ superalgebra representations.
To make our exposition self-contained we employ some notation and results in connection
with a continuum $su(2|1)$ path integral already reported elsewhere \cite{k2}.  However, the conclusion we reach below
requires a more sophisticated approach based on
a carefully defined time-lattice representation of the corresponding path integral.

Just to get an idea about  what strong correlations really are, consider
the $U=\infty$ Hubbard model which is known to capture the extreme limit for strongly correlated electrons.
The Hamiltonian reads
\begin{equation}
H_t=-\sum_{ij,\sigma}t_{ij}c^{\dagger}_{i\sigma}c_{j\sigma}-\mu\sum_{i}(1-n_i),
\quad n_i=\sum_{\sigma}n_{i\sigma}\le 1.
\label{1.01}\end{equation}
Here $c^{\dagger}_{i\sigma} (c_{i\sigma})$ is an on-site creation (annihilation) operator of an
electron excitation with the spin projection $\sigma=\uparrow, \downarrow$.
The hopping amplitudes $t_{ij}$ represent jumps between
nearest-neighbor (nn) and next-to-nearest-neighbor (nnn)sites and are zero otherwise.
Throughout this paper we will be considering a $D$ dimensional bipartite lattice,
$L=A\oplus B.$
The chemical potential $\mu$ controls
the total number of vacancies (empty sites).

The infinitely large on-site Coulomb repulsion is accounted for
by the local NDO constraint, $n_i\le 1$, that restricts the on-site Hilbert space to states with at most one electron per site.
This is the essence of strong electron correlations.
In the absence of this constraint, this  model simply  describes a system of noninteracting electrons
and it can be trivially diagonalized in the momentum space.
The NDO constraint makes this problem nontrivial and its exact solution is still unknown for spatial dimensions $>1D$.
The physics behind the model (\ref{1.01}) is certainly far from trivial.
Indeed, one of the few exact results was proved by Nagaoka \cite{nagaoka} who showed that for one hole
the ground state of the $U=\infty$ Hubbard model is a fully saturated
ferromagnet. This provides an interesting example of a quantum system
with ferromagnet ordering due to a purely kinetic-energy effect driven by
hole hopping (itinerant ferromagnetism).
Unfortunately, despite very extensive work over the years, both this model and
itinerant ferromagnetism are still poorly understood.
One of the open important questions related to this concerns the
thermodynamic stability of the Nagaoka phase.
That is, whether or not the Nagaoka state is stable when
the density of holes is finite in the thermodynamic limit.

The local NDO constraint can be explicitly incorporated into the theory by
projecting the Hamiltonian (\ref{1.01}) onto the restricted Hilbert space$\prod_i {\cal H}^{phys}_i,$ where
the $3D$ on-site Hilbert space ${\cal H}^{phys}_i$ is
spanned by the vectors $|0\rangle_i$ (empty site), $|\uparrow\rangle_i$
(spin-up electron), and $|\downarrow\rangle_i$ (spin-down electron):
$$H\to H={\cal P}H{\cal P},\quad {\cal P}=\prod_i{\cal P}_i,$$ where the Gutzwiller projection operator ${\cal P}_i$
is given by
$${\cal P}_i=1-n_{i\sigma}n_{i-\sigma},$$ so that
$${\cal P}_ic_{i\sigma}{\cal P}_i=:\tilde c_{i\sigma}=c_{i\sigma}(1-n_{i-\sigma}).$$
In this $3D$ subspace the constrained electron operators $\tilde c_{i\sigma}$ can be identified with the
Hubbard operators,
$$ X^{0\sigma}=(X^{\sigma 0})^{\dagger}=|0\rangle\langle\sigma|,
\quad \sigma=\uparrow,\downarrow.$$ It follows that
$\tilde c_{i\sigma}=X_i^{0\sigma},$
provided $n_{i\sigma}n_{i-\sigma}=0.$ The last requirement eliminates the doubly occupied fermionic states.
As a result the Hamiltonian (\ref{1.01}) in the constrained physical space takes the form
\begin{eqnarray}
H_t&=&-t\sum_{ij,\sigma}\tilde c^{\dagger}_{i\sigma}\tilde c_{j\sigma}-\mu\sum_{i}(1-\sum_{\sigma}\tilde c^{\dagger}_
{i\sigma}
\tilde c_{i\sigma})\nonumber\\
&=&-t\sum_{ij,\sigma}X_i^{\sigma 0}X_j^{0\sigma}-\mu\sum_iX^{00}_i.
\label{1.1}\end{eqnarray}
Here $X^{00}=X^{0\sigma}X^{\sigma 0}=|0\rangle\langle 0|=1-\sum_{\sigma}\tilde c^{\dagger}_{\sigma}
\tilde c_{\sigma}$ stands for the on-site vacancy number operator, so that the concentration of vacancies becomes
$\delta=\frac{1}{N_s}\sum_i\langle X^{00}_i\rangle.$
The on-site operator $X^{0\sigma}$ removes
an electron with the spin projection $\sigma$ and creates  a vacancy (empty state) which is a spin singlet.

The important point is that the fermionic Hubbard operators
$X^{0\sigma}=(X^{\sigma 0})^{\dagger}$ along with the bosonic ones, $X^{\sigma\sigma'},\, X^{00},$ are closed under
commutation/anticommutation relations into the {\it superalgebra} $su(2|1)$ \cite{wiegmann}.
The $su(2|1)$ superalgebra can be thought of as the simplest possible extension of the conventional
spin $su(2)$ algebra to incorporate fermionic degrees of freedom.
Namely, the  bosonic sector of the $su(2|1)$ consists of three  bosonic superspin operators,
\begin{equation}
Q^{+}=X^{\uparrow\downarrow},\quad Q^{-}=X^{\downarrow\uparrow},\quad
Q^{z}=\frac{1}{2}(X^{\uparrow\uparrow}-X^{\downarrow\downarrow})
\label{qx}\end{equation}
closed into $su(2)$, and a
bosonic operator $X^{00}$ that generates a $u(1)$ factor of the maximal even subalgebra $su(2)\times u(1)$
of $su(2|1)$. The fermionic sector is constructed out of four operators $X^{\sigma 0}, X^{0\sigma}$
that transform in a spinor representation of $su(2)$.
We do  not intend here to discuss a general theory of the $su(2|1)$ irreducible
representations that can be found elsewhere \cite{sch}.
We will focus instead on a specific {\it lowest} $3D$ irreducible representation of $su(2|1)$
generated by the Hubbard operators.
This representation acts in the physical
on-site Hilbert space of model (\ref{1.01}) with the basis vectors, $|\uparrow\rangle, |\downarrow\rangle, |0\rangle$,  and
consists of nine operators $X^{\lambda\lambda'}=|\lambda\rangle\langle \lambda'|,\quad \lambda,\lambda'=0,\sigma.$
The resolution of unity in this space,
\begin{equation}
\hat I=|0\rangle\langle 0|+|\uparrow\rangle\langle\uparrow|+|\downarrow\rangle\langle\downarrow|=
\sum_{\lambda}X^{\lambda\lambda},
\label{space}\end{equation}
singles out eight independent generators of $su(2|1)$ listed above.
The special important property of this representation is that
$X^{\lambda\lambda'}X^{\lambda''\lambda'''}=\delta_{\lambda'\lambda''}X^{\lambda\lambda'''}$.

\section{$su(2|1)$ coherent-state manifold}

The normalizable coherent states (CS's) associated with the lowest irreducible representation of $su(2|1)$
superalgebra spanned by Hubbard operators take the form
\begin{eqnarray}
|z,\xi\rangle&=&(1+\bar{z}z
+\bar{\xi}\xi)^{-1/2}\exp\left(zX^{\downarrow\uparrow}+\xi
X^{0\uparrow}\right)|\uparrow\rangle \nonumber\\
&=&(1+\bar{z}z +\bar{\xi}\xi)^{-1/2}(|\uparrow\rangle
+z|\downarrow\rangle+\xi |0\rangle), \label{1.2}
\end{eqnarray} where a complex even Grassmann parameter $z$ and  an odd complex
Grassmann parameter $\xi$ are the inhomogeneous (proper) coordinated of a point on a supersphere,
$(z,\xi) \in S^{2|2}\simeq CP^{1|1}=SU(2|1)/U(1|1).$ Here $CP^{1|1}$ stands for a complex
projective superspace with a complex
dimension $(1,1)$. It can be thought of as a minimal superextension of an ordinary projective space $CP^1$ homeomorphic
to a two-sphere, $CP^1\simeq S^2$.
The odd Grassmann parameter appears in (\ref{1.2}) due to
the fact that $X^{\downarrow 0}$ is a fermionic operator in
contrast with the operator $X^{\downarrow\uparrow}$. The product
$\xi X^{0\uparrow}$ represents therefore a bosonic quantity as
required.

One can use as well the homogeneous coordinates $(z^1,z^2,\theta)$ on the supersphere, so that $z=z^1/z^2,\xi=\theta/z^2,$
with $z^2\neq 0$.
Then the coset $CP^{1|1}$ manifold is defined by the equation $$|z^1|^2+|z^2|^2+\bar\theta\theta=1.$$
The supergroup $SU(2|1)$ acts on $CP^{1|1}$ according to
$Z\to Z^g=gZ,$ where$Z=(z^1,z^2,\theta)^t$ and $g\in SU(2|1).$ This generates a corresponding transformation of
the inhomogeneous coordinates, $(z,\xi)$.
In particular, if one chooses $g$ to represent a pure spin rotation,
$$g=
 \left(\begin{array}{lll}
u & v&0 \\
-\overline{v} & \overline{u}&0\\
0&0&1
\end{array}
\right),\quad
 \left(\begin{array}{ll}
u & v \\
-\overline{v} & \overline{u}%
\end{array}
\right) \in \mathrm{SU(2)},
$$
one gets
\begin{equation}
z\to \frac{uz+v}{-\overline{v}z+\overline{u}},\quad \xi\to \frac{\xi}{-\overline{v}z+\overline{u}}.
\label{2field}\end{equation}
Note that both the bosonic and fermionic fields transform themselves under $SU(2)$ spin rotations.

At $\xi =0$, the su$(2|1)$ CS reduces to the ordinary spin su(2) CS,
\begin{equation}
|z,\xi=0\rangle= |z\rangle_{s=1/2}\equiv |z\rangle=\frac{1}{\sqrt{1+|z|^2}}\exp( zS^{-})|\uparrow\rangle=
\frac{1}{\sqrt{1+|z|^2}}(|\uparrow\rangle +z|\downarrow\rangle),
\label{1.2a}\end{equation}
where the complex number $z$
is a stereographic coordinate of a point on an ordinary sphere, $z\in S^2\simeq CP^1=SU(2)/U(1)$.
The spin operators $\vec S$ obey the standard commutation relations
\begin{equation}
[S_z,S_{\pm}]=\pm S_{\pm},\quad [S_{+},S_{-}]=2S_z, \quad \vec
S^2=3/4.\label{1.2b}\end{equation}
These operators coincide with bosonic generators $\vec Q,$ of $su(2|1)$ at half filling,
in which case the on-site Hilbert space is reduced and spanned only by the vectors
$|\uparrow\rangle, \,|\downarrow\rangle$.
In contrast, at $z=0$, the state $|\xi\rangle\equiv |z=0,\xi\rangle$ represents a pure fermionic CS.

At this stage it is helpful to introduce the important notion of the covariant (Berezin) symbol for
a Hubbard operator $X$.
It can also be referred to as a coherent-state symbol and is defined as follows
\begin{equation}
X_{cov}:=\langle z,\xi|X|z,\xi\rangle.
\label{symbol}\end{equation}
Explicitly, we find
\begin{eqnarray}
X^{0\downarrow}_{cov}&=&-\frac{z\bar\xi}{1+|z|^2},\quad
X^{\downarrow 0}_{cov}=-\frac{\bar
z\xi}{1+|z|^2},\nonumber\\
X^{0\uparrow}_{cov}&=&-\frac{\bar\xi}{1+|z|^2}, \quad X^{\uparrow
0}_{cov}=-\frac{
\xi}{1+|z|^2},
\label{x}\end{eqnarray}
\begin{eqnarray}
Q^{+}_{cov}=S^{+}_{cov}\left(1-X^{00}_{cov}\right),\quad Q^{-}_{cov}=S^{-}_{cov}
\left(1-X^{00}_{cov}\right),\quad
Q^z_{cov}=S^z_{cov}
\left(1-X^{00}_{cov}\right),
\label{1.3}\end{eqnarray}
where the covariant symbol of the hole number operator reads
\begin{equation}
X_{cov}^{00}=\frac{\bar\xi\xi}{1+|z|^2}.
\label{1.3a}\end{equation}
The corresponding CS symbols of the su(2) generators are evaluated
to be ($S_{cov}:=\langle z|S|z\rangle$):
\begin{eqnarray}
S^{+}_{cov}=\frac{z}{1+|z|^2},\quad
S^{-}_{cov}=\frac{\bar z}{1+|z|^2},\quad
S^z_{cov}=\frac{1}{2}\left(\frac{1-|z|^2}{1+|z|^2}\right).
\label{1.3b}\end{eqnarray}
The important point is that the covariant symbol of the Hamiltonian associated with
the algebra generators enters the path-integral action for the
partition function.
For a compact simple algebra those symbols are in one-to-one
correspondence with the algebra generators. This is the case for both the $su(2|1)$ and $su(2)$ algebras.
Note also that
at half filling, $\delta=0$, we get $X_{cov}^{00}=0$ so that the symbol of the $\vec Q$ operator
reduces to that of the conventional spin operator, $\vec Q_{cov}=\vec S_{cov}$. In general this is not the case.
Although the generalized
spin operators $\vec Q$ fulfill the same commutation relations as operators $\vec S$ do, the
operator $\vec Q^2$ is not a $c$-number. Namely, $\vec Q^2=\vec S^2(1+X^{00}).$ This means that operators $\vec Q$ are
closed into an algebra larger than $su(2)$.

The $su(2|1)$ CS is parameterized by the coordinates of a point on
the coherent-state supermanifold $CP^{1|1}$. The latter  appears as a classical phase space
for the constrained electrons. The symplectic structure is
given
by a canonical symplectic two-form,
\begin{equation}
\Omega=d(i\langle z,\xi|d|z,\xi\rangle)=-i\left[dz(a)d\bar z+dz(\alpha)d\bar\xi+d\xi(\gamma)d\bar z+d\xi(y)d\bar\xi\right],
\label{s1}\end{equation}
where the external derivative
$$d=dz\frac{\partial}{\partial z}+d\bar z\frac{\partial}{\partial \bar z}+d\bar\xi\frac{\partial}
{\partial \bar\xi}+d\xi\frac{\partial}{\partial \xi}$$ and
$$a=\frac{1+\bar\xi\xi}{(1+|z|^2+\bar\xi\xi)^2},\, \alpha=\frac{\bar z\xi}{(1+|z|^2+\bar\xi\xi)^2},\, \gamma=
\frac{z\bar\xi}{(1+|z|^2+\bar\xi\xi)^2},\, y=-\frac{1}{1+|z|^2}.$$
We use the convention that a superform on $CP^{1|1}$ is $Z\times Z_2$ graded, where the $Z$- gradation is the usual
gradation of the de Rham complexes, while the $Z_2$-gradation is a natural gradation of Grassmann algebra \cite{gr}.
We thus have
$$dzd\bar z=-d\bar zdz,\quad dzd\bar\xi=-d\bar\xi dz,\quad d\xi d\bar\xi=d\bar\xi d\xi.$$

Classical dynamics on $CP^{1|1}$ is governed by the Poisson brackets generated by the symplectic structure.
Namely, for any two functions $g$ and $h$ that represent classical observables on $CP^{1|1}$, we get
\begin{equation}
\{g,h\}_{PB}=+iX_g\rfloor dh,
\label{s2}\end{equation} where $X_g$ denotes a vector field associated to $g$ and the symbol $\rfloor$ stands for the
interior
product (or contraction), so that \cite{nakahara}
$$X_g\rfloor \Omega=dg.$$ In particular, the covariant symbols (\ref{x},\ref{1.3}) are closed into the $su(2|1)$
superalgebra under the
action of brackets (\ref{s2}).
Up to a multiplicative constant the $SU(2|1)$ invariant measure is given by
\begin{equation}
d\mu\propto sdet ||\Omega_{a\bar b}||d\bar zdzd\bar\xi d\xi\propto \frac{d\bar zdzd\bar{\xi} d\xi}{1+\bar z z+\bar{\xi}\xi},
\label{s3}\end{equation}
where $"sdet"$ stands for the superdeterminant (or Berezian) \cite{berezin}, while $ ||\Omega_{a\bar b}||$ stands for the
supermatrix form of $\Omega$, namely $\Omega=dx^a\Omega_{a\bar b}dx^{\bar b},\,\,x^a=z,\xi.$

\section{$su(2|1)$ path integral: preliminaries}

We now seek a path-integral representation of the partition function $Z=tr e^{-\beta H}$
where $H$ is expressed in terms of Hubbard operators. This implies that the NDO constraint
is explicitly resolved from the outset,
so that no redundant gauge dependent variables emerge within that approach.
To work out the path-integral formalism to deal directly with  the Hubbard operators,
consider the on-site $3D$ physical Hilbert space ${\cal H}_{phys}$ spanned by the vectors $|\lambda\rangle,
\,\lambda=\uparrow,\downarrow,0.$ The Hubbard operators act in this space and are naturally $Z_2$ graded. As was already
mentioned, there are bosonic
or even-graded and fermionic or odd-graded operators. Accordingly, the basis in ${\cal H}_{phys}$ becomes graded
as well. We choose the spin-up $|\uparrow\rangle$ and spin-down $|\downarrow\rangle$
states to be even-graded, whereas the empty state (vacancy) $|0\rangle$
is considered to be odd-graded. This means that odd Grassmann parameters commute with the spin states and anti-commute
with the vacancy vector. From now on we consider the grading of the physical $3D$ subspace to be fixed in
accordance with this.
For any operator $H$ acting in ${\cal H}_{phys}$ one then gets
\begin{equation}
tr e^{-\beta H}=\sum_{\lambda}\langle\lambda|e^{-\beta H}|\lambda\rangle,
\label{1.5}\end{equation}
and
\begin{equation}
str e^{-\beta H}=\sum_{\lambda}(-)^{grad(|\lambda\rangle)}\langle\lambda|e^{-\beta H}|\lambda\rangle,
\label{1.55}\end{equation}
By using the $su(2|1)$ CS (\ref{1.2}), Eq.(\ref{1.5}) can be rewritten in the form
\begin{equation}
tr e^{-\beta H}=\int d\mu\langle z,\xi|e^{-\beta H}|z,-\xi\rangle.
\label{1.6}\end{equation}
Here the $SU(2|1)$ invariant measure
\begin{equation}
d\mu_{su(2|1)}\equiv d\mu=\frac{d\bar zdzd\bar{\xi} d\xi}{2\pi i(1+\bar z z+\bar{\xi}\xi)}
\label{1.7}\end{equation}
agrees with Eq. (\ref{s3}). Integration over $\bar z,z$ is performed on a complex plane. The integration over
Grassmann parameters is carried out according to the standard rules \cite{berezin},
$$\int d\xi=\int d\bar\xi=0,\quad \int d\xi \xi=\int d\bar\xi\bar\xi=1.$$
Notice that in contrast with Eq.(\ref{1.6})
\begin{equation}
str e^{-\beta H}=\int d\mu\langle z,\xi|e^{-\beta H}|z,\xi\rangle.
\label{1.66}\end{equation}
The distinction between Eqs. (\ref{1.6}) and (\ref{1.66})
results in different boundary conditions (anti-periodic versus periodic)
for Grassmann amplitudes in the corresponding path integrals for the partition/superpartition functions.

In the course of the derivation
of the partition function we essentially rely on a certain change of the path-integral variables.
To justify it,
we start with an example of a
simple single-site Hamiltonian that also admits a treatment in terms of ordinary
multiple integrals.
After that, we reconsider the problem in terms of the time-continuous path integral.

Consider the single-site Hamiltonian
\begin{equation}
H=\mu X^{00}.
\label{single}\end{equation}
Operator $X^{00}=|0\rangle\langle 0|$ is represented in ${\cal H}_{phys}$ by $3\times 3$ matrix
with eigenvalues $0,\, 0,\, 1$. As a result, the partition function reduces to
\begin{equation}
Z_0=2+e^{-\beta \mu},
\label{mu}\end{equation}
and the vacancy occupation number becomes
\begin{equation}
\delta=\langle X^{00}\rangle=-\frac{1}{\beta}\partial_{\mu}\log Z_0=\frac{1}{2e^{\beta\mu}+1}.
\label{delta}\end{equation}
This is to be compared with the conventional spinless fermion occupation number,
$$n_F=\frac{1}{e^{\beta\mu}+1}.$$

Since $(X^{00})^2=X^{00}$ we get
$$e^{-\mu\beta X^{00}}=1+X^{00}(e^{-\mu\beta}-1).$$
Equation (\ref{1.6}) then gives
\begin{eqnarray}
tr e^{-\beta\mu X^{00}}&=&\int d\mu\langle z,\xi|(1+X^{00}(e^{-\mu\beta}-1))|z,-\xi\rangle\nonumber\\
&=& \int\frac{d\bar zdzd\bar\xi d\xi}{2\pi i(1+\bar zz+\bar\xi\xi)}\left[\frac{1+\bar zz-\bar\xi\xi}{1+\bar zz+\bar\xi\xi}
-\frac{\bar\xi\xi}{1+\bar zz+\bar\xi\xi}(e^{-\beta\mu}-1)\right].
\label{a}\end{eqnarray}
Here we have used that
\begin{equation}
\langle z',\xi'|z,\xi\rangle=\frac{1+\bar z'z+\bar\xi'\xi}{\sqrt{1+\bar z'z'+\bar\xi'\xi'}
\sqrt{1+\bar zz+\bar\xi\xi}}.
\label{b}\end{equation}

Although integral (\ref{a}) can be calculated immediately, it is instructive at this stage to make
a change of variables
\begin{equation}
z\to z,\quad \xi\to\xi\sqrt{1+|z|^2}
\label{change}\end{equation}
to bring the measure into a more tractable form,
\begin{equation}
d\mu=\frac{d\bar zdzd\bar\xi d\xi}{2\pi i(1+\bar zz+\bar\xi\xi)}\to
\frac{d\bar zdzd\bar\xi d\xi}{2\pi i(1+\bar zz)^2}(1-\bar\xi\xi).
\label{changem}\end{equation}
Equation (\ref{a}) then becomes
\begin{eqnarray}
tr e^{-\beta\mu X^{00}}
&=& \int\frac{d\bar zdzd\bar\xi d\xi}{2\pi i(1+\bar zz)^2}(1-\bar\xi\xi)
\left[\frac{1-\bar\xi\xi}{1+\bar\xi\xi}
-\bar\xi\xi(e^{-\beta\mu}-1)\right]\nonumber\\
&=&\int\frac{d\bar zdzd\bar\xi d\xi}{2\pi i(1+\bar zz)^2}\left[1-3\bar\xi\xi -\bar\xi\xi(e^{-\beta\mu}-1)\right])
=3+e^{-\beta\mu}-1=2+e^{-\beta\mu},
\label{b}\end{eqnarray}
which agrees with (\ref{mu}).

Under the change (\ref{change}) we get $X^{00}_{cov}\to \bar\xi\xi,$
so that it may seem we have reduced the problem to that of spinless fermions. However,
this is not the case.
The measure (\ref{changem}) involves the extra factor $(1-\bar\xi\xi)$
comparing with the standard product of the $SU(2)$ invariant spin and spinless fermion measures,
$$\frac{d\bar zdz}{2\pi i(1+\bar zz)^2}d\bar\xi d\xi.$$
This extra factor reflects a nontrivial symplectic structure of the underlying path-integral
phase space. Namely, the transformation (\ref{change}) brings the symplectic two-form in
\begin{equation}
\Omega\to \bar\Omega=-i\left[dz(\bar a)d\bar z+dz(\bar\alpha)d\bar\xi+d\xi(\bar\gamma)d\bar z+d\xi(\bar y)d\bar\xi\right],
\label{s4}\end{equation}
where
$$\bar a=\frac{1-\bar\xi\xi+\bar\xi\xi |z|^2}{(1+|z|^2)^2},\, \bar\alpha=\frac{\bar z\xi}{1+|z|^2},\, \bar\gamma=
\frac{z\bar\xi}{1+|z|^2},\, \bar y=-1.$$
It then follows immediately that
$$d\mu\to d\bar\mu\propto sdet ||\bar\Omega_{a\bar b}||d\bar zdzd\bar\xi d\xi\propto \frac{1-\bar\xi\xi}{(1+|z|^2)^2}
d\bar zdzd\bar\xi d\xi ,$$
which agrees with Eq.( \ref{changem}).
Poisson brackets with respect to (\ref{s4}) then give us
$$\{\xi,\bar\xi\}_{PB}=1-\bar\xi\xi|z|^2,\quad \{\xi,\vec S_{cov}\}_{PB}\ne 0,$$
which clearly implies that the amplitudes $\xi,\bar\xi$ do not represent spinless fermions
independent of lattice spins: vacancies and lattice spins are correlated due to
the NDO constraint. For example, the destruction of a vacancy necessarily results in the creation
of a lattice spin.
As we see below,  a calculation of a purely fermionic
correlater  involves both the $\xi$ and $z$ variables in a nontrivial manner.

If we ignored the factor $(1-\bar\xi\xi)$ in the measure  we would end up with
a partition function of a spinless fermion,
$Z=1+e^{-\beta\mu}$.
The conventional fermionic amplitudes $f,\bar f$ obey
the conventional rules,  $\{f,\bar f\}_{PB}=1, \quad  \{f,\vec S_{cov}\}_{PB}= 0,$
with the symplectic structure taking on the standard form,
$$\Omega \propto \frac{dzd\bar z}{(1+|z|^2)^2}-df d\bar f.$$

The composite field $(z,\xi)$ parameterizes a point on a supersphere, $S^{2|2}\simeq CP^{1|1}.$
For the conventional spinless fermions coupled to $su(2)$ spins
the underlying phase space is
instead given by a direct product of the ordinary sphere, $S^2\simeq CP^1$, and a complex Grassmann plane.

\section{$su(2|1)$ path integral: partition function}

We turn now to a derivation of the path-integral representation of the partition function.
A key ingredient in constructing the CS path integral is the resolution of unity in ${\cal H}_{phys}$,
\begin{equation}
I=\int d\mu|z,\xi\rangle\langle z,\xi|,
\label{1.8}\end{equation}
where the measure is given by (\ref{1.7}).
This equation can be used repeatedly to compute $\langle z,\xi|e^{-\beta H}|z,-\xi\rangle$,
considering $H$ to be a local on-site Hamiltonian expressible in terms of the $su(2|1)$ generators.
Following standard procedure, let us  break up the interval $[0,\beta]$ into $N$ small pieces of length
$\epsilon=\beta/N,\, N\to\infty$.
Then Eq.(\ref{1.6}) can be rewritten in the form
\begin{equation}
tr e^{-\beta H}=\int d\mu \int \prod_{k=0}^{N}d\mu_k\langle z,\xi|N\rangle\langle N|N-1\rangle\cdots\langle 0|z,-\xi\rangle
e^{-\epsilon \sum_k H(k,k-1)} +{\cal O}(\epsilon^2).
\label{1.9}\end{equation}
Here
\begin{equation}
H(k,k-1)=\frac{\langle k|H|k-1\rangle}{\langle k|k-1\rangle},\quad |k\rangle:=|z_k,\xi_k\rangle,
\quad z_k=z(\epsilon k),\,\xi_k=\xi(\epsilon k),
\quad k=1,...,N,
\label{mar}\end{equation}
and equation (\ref{1.6}) tells us that
$$\int d\mu\langle z,\xi|N\rangle\langle 0|z,-\xi\rangle= tr |N\rangle\langle 0|=\langle z_0,\xi_0|z_N,-\xi_N\rangle.$$
Finally, integrating over $d\mu$ in (\ref{1.9}) yields
\begin{equation}
tr e^{-\beta H}=\int d\mu_0\prod_{k=1}^{N}d\mu_k\langle k|k-1\rangle\langle z_0,\xi_0|z_N,-\xi_N\rangle
e^{-\epsilon \sum_k H(k,k-1)}.
\label{1.10}\end{equation}
From now on we drop the ${\cal O}(\epsilon^2)$ contribution to the partition function having in mind
that the continuum limit will be taken eventually.
We further notice that the kernel $\langle z,\xi|z',\xi'\rangle$
acts as a delta function with respect to measure $d\mu$. To see this consider any vector $|\psi\rangle$ that
belongs to ${\cal H}_{phys}$. Then resolution of unity implies that
$$\langle\psi|=\int d\mu\langle\psi|z,\xi\rangle\langle z,\xi|,$$
which in components reads simply
$$\psi(z',\xi')=\int d\mu \psi(z,\xi)\langle z,\xi|z',\xi'\rangle,\quad \psi(z,\xi):=\langle
\psi|z,\xi\rangle.$$
Having this in mind, Eq.(\ref{1.10}) becomes
\begin{eqnarray}
tr e^{-\beta H}&=&\int\prod_{k=1}^{N}d\mu_k\langle k|k-1\rangle
e^{-\epsilon \sum_k H(k,k-1)}\mid_{z_0=z_N,\,\xi_0=-\xi_N}\nonumber\\
&=&
\int \prod_{k=1}^Nd\mu_k\exp{\sum_{k=1}^N\left[\log\langle k|k-1\rangle
-\epsilon H(k,k-1)\right]}\mid_{z_0=z_N,\,\xi_0=-\xi_N}
\label{1.11}\end{eqnarray}
Performing here a formal time-continuum limit $\epsilon\to 0$ yields
\begin{eqnarray}
tr e^{-\beta H}&=&\int D\mu e^{\int_0^{\beta}Ld\tau},
\label{1.110}\end{eqnarray}
where
\begin{eqnarray}
L=-\langle z,\xi|\frac{\partial}{\partial \tau}+H|z,\xi\rangle
\label{1.111}\end{eqnarray}
and
\begin{eqnarray}
D\mu_{su(2|1)}(z,\xi)\equiv D\mu(z,\xi)=
\prod_{\tau}\frac{d\bar z(\tau)dz(\tau)d\bar{\xi}(\tau)d\xi(\tau)}{2\pi i(1+\bar z z+\bar{\xi}\xi)}.
\label{1.112}\end{eqnarray}
The $SU(2|1)$ symplectic potential explicitly reads
\begin{eqnarray}
\langle z,\xi|-\frac{\partial}{\partial \tau}|z,\xi\rangle=
\frac{1}{2}\left(\frac{{\dot{\bar z}} z-\bar z\dot z+{\dot{\bar\xi}}\xi-\bar\xi\dot\xi}{1+|z|^2+\bar\xi\xi}\right),
\label{1.113}\end{eqnarray}
and $\langle z,\xi|H|z,\xi\rangle=H_{cov}.$

We now specify
Hamiltonian to be that of the $U=\infty$ Hubbard model given by Eq.(\ref{1.1})
and make the change of variables (\ref{change}). We then get
\begin{eqnarray}
tr e^{-\beta H}&=&\int\prod_i\prod_{k=1}^{N}\frac{d\bar z_i(\epsilon k)dz_i(\epsilon k)
d\bar\xi_i(\epsilon k) d\xi_i(\epsilon k)}{2\pi i(1+\bar z_i(\epsilon k)z_i(\epsilon k))^2}(1-\bar\xi_i(\epsilon k)
\xi_i(\epsilon k))\nonumber\\
&\times&
\exp{\sum_{k=1}^N\left[ \sum_i\log A_i(k)  -\epsilon H_{cov}(k,k-1)\right]},\nonumber\\
\label{1.12}\end{eqnarray}
accompanied with the boundary conditions $z_0=z_N,\,\xi_0=-\xi_N$.
Here
\begin{equation}
H_{cov}(k,k-1)=t\sum_{ij}\bar\xi_j(\epsilon k)\xi_i(\epsilon (k-1))\langle z_i(\epsilon k)|z_j(\epsilon (k-1)
\rangle-\mu\sum_i
\bar\xi_i(\epsilon k)\xi_i(\epsilon (k-1)),
\label{1.13}\end{equation}
and
$$A_i(k) =\frac{1+\bar z_{i}(\epsilon k)z_{i}(\epsilon (k-1)}
{\sqrt{(1+|z_{i}(\epsilon k)|^2)(1+|z_i(\epsilon (k-1))|^2)}}
(1-\frac{1}{2}\bar\xi_{i}(\epsilon k)\xi_{i}(\epsilon k))
(1-\frac{1}{2}\bar\xi_i(\epsilon(k-1))\xi_i (\epsilon(k-1)))
+\bar\xi_i(\epsilon k)\xi_i(\epsilon(k-1)).$$
Let us recall that the indices $i,j$ denote the lattice sites, whereas the index $k$ numerates
time slices only.

In the continuum $\epsilon\to 0$ limit, Eq.(\ref{1.12}) becomes
\begin{equation}
tr e^{-\beta H}=\int D\mu (z,\xi )\
e^{\int_0^{\beta}L(z, \xi)d\tau}, \label{1.14}
\end{equation}
where
\begin{equation}
D\mu (z,\xi )=\prod_{i,\tau}\frac{d\bar z_i(\tau)dz_i(\tau)d\bar\xi_i(\tau)d\xi_i(\tau)}{2\pi i(1+|z_i|^2)^2}\,
(1-\bar\xi_i\xi_i)
\label{1.15} \end{equation} stands for
the measure with the boundary conditions,
$z_i(0)=z_i(\beta),\, \xi_i(0)= -\xi_i(\beta).$ The Lagrangian now reads
\begin{eqnarray}
L&=&\sum_i ia^{(0)}_i(\tau)-\sum_i\bar\xi_i
\left(\partial_{\tau}+\mu+ia^{(0)}_i\right)\xi_i-
H_{cov}.
\label{1.16}\end{eqnarray}
The first piece of the action  involves
the time component of the Berry connection to be discussed later,
$$ia^{(0)}=-\langle z|\partial_{\tau}|z\rangle=\frac{1}{2}\frac{\dot{\bar z}z-\bar z\dot
z}{1+|z|^2}.$$
In the time-discretized representation, it reads
\begin{equation}
ia^{(0)}(\epsilon k)\equiv ia_k=\frac{1}{2}\frac{\bar\delta_k z_k-\delta_k \bar z_k}
{1+|z_k|^2}=\log \langle z_k|z_{k-1}\rangle +{\cal O}(\delta_k^2),
\quad \delta_k=z_k-z_{k-1}.
\label{berryd}\end{equation}

The dynamical part of the action is given by
\begin{eqnarray}
H_{cov}=
t\sum_{i\ne j}\overline{\xi }_{j}\xi _{i}\langle
z_{i}|z_{j}\rangle -\mu\sum_i\bar\xi_i\xi_i.
\label{1.17}
\end{eqnarray}
Here $\langle z_{i}|z_{j}\rangle$ stands for the product of the spin
coherent states,
\begin{equation}
\langle z_{i}|z_{j}\rangle =\frac{1+\overline{z}
_{i}z_{j}}{\sqrt{(1+|z_{j}|^{2})(1+|z_{i}|^{2})}}.
\label{cs}\end{equation}
The covariant symbol of the on-site electron spin operator reduces to
\begin{equation}
\vec Q^{cov}_i=
\vec S_i^{cov}(1-\bar\xi_i\xi_i).\label{symbolQ}\end{equation}

Now we move on to a practical calculation of path integral (\ref{1.14}).
However, to be more accurate we do that right on the time lattice.
As we see below, that will enables us to clarify some subtle points concerning the structure of
the path-integral action which is not seen in the naive continuum limit.

The major problem is of course the factor
$\prod_k(1-\bar\xi_k\xi_k)$ presented in the on-site measure. Since one can rewrite this factor
as
$$\exp{\left(-\sum_k\bar\xi_k\xi_k\right)}=\exp{\left(-\frac{1}{\epsilon}\sum_k\bar\xi_k\xi_k\epsilon\right)}\to
\exp{\left(-\frac{1}{\epsilon}\int_0^{\beta}\bar\xi\xi d\tau\right)},\quad\epsilon=\beta/N\to 0,\quad N\to\infty$$
it might seem that in the continuum limit it simply amounts to an additive renormalization of the
chemical potential and as such can be discarded \cite{karchev}. However, this is not the case.
To start with, the renormalization is infinite which makes
the whole procedure rather formal. Besides, in the chemical potential term from representation
(\ref{1.13})
the arguments of the $\bar\xi$ and $\xi$ fields are shifted by one  time step, whereas the factor in the measure
involves the
$\bar\xi, \,\xi$ fields at  coinciding time moments. This slight difference happens to affect
the final results drastically.

To see this, consider path integral (\ref{1.14}) with the action
\begin{eqnarray}
S&=&\int_0^{\beta}ia^{(0)}(\tau)d\tau-\int_0^{\beta}\bar\xi
\left(\partial_{\tau}+\mu+ia^{(0)}\right)\xi d\tau.
\label{1.20}\end{eqnarray}
With this action the path-integral evaluation of the partition function should reproduce the earlier result
$$Z=tr e^{-\beta\mu X^{00}}=2+e^{-\beta\mu}$$
which we already calculated in Eq.(\ref{b}) using ordinary integrals.
By a $U(1)$ phase transformation of the fermionic fields  the potential $a^{(0)}(\tau)$ can be brought into a time
independent form,
$$a^{(0)}\to a^{(0)}-\dot\phi=\frac{1}{\beta}\int_0^{\beta}ad\tau,$$ where
\begin{equation}
\phi(\tau)=-\frac{\tau}{\beta}\int_0^{\beta}a^{(0)}ds +\int_0^{\tau}a^{(0)}ds.
\label{phase}\end{equation}
Note that $\phi(0)=\phi(\beta)$.
The effective action then becomes
\begin{equation}
S=\int_0^{\beta}ia^{(0)}(\tau)d\tau-\int_0^{\beta}\bar\xi
\left(\partial_{\tau}+\bar{\mu} \right)\xi d\tau,
\label{1.21}\end{equation}
where $\bar{\mu}=\mu+\frac{1}{\beta}\int_0^{\beta}ia^{(0)}d\tau.$
Here we cannot simply
integrate out the $\xi$ fields in (\ref{1.21}) by standard means,
since now the measure contains the extra factor, $\exp{\left(-\sum_k\bar\xi_k\xi_k\right)}$.
To figure out how the path integral works in this case we must look back at the defining time-discretized
representation (\ref{1.12}). It becomes
\begin{eqnarray}
tr e^{-\beta\mu X^{00}}=\int\prod_{k=1}^{N}\frac{d\bar z_kdz_k
d\bar\xi_kd\xi_k e^{ia_k}}{2\pi i(1+\bar z_kz_k)^2}e^{-S},
\label{1.22}\end{eqnarray}
where
\begin{eqnarray}
S=\epsilon\sum_{k=2}^N\bar\xi_k \left(\frac{(\xi_k-\xi_{k-1})}{\epsilon}+\overline\mu\xi_{k-1}\right)
+\epsilon \bar\xi_1\left(\frac{(\xi_1+\xi_N)}{\epsilon}-\overline\mu\xi_{N}\right)+\sum_{k=1}^N\bar\xi_k \xi_k,
\quad \xi_k=\xi(\epsilon k).
\label{1.23}\end{eqnarray}
Here we have taken into account that $\xi_0=-\xi_N$. The last term in the action represents the contribution
that originates from the extra measure factor, so that the integral over the fermionic fields reduces
simply to
$$\lim_{N\to\infty}\int\prod_{k=1}^{N}d\bar\xi_kd\xi_ke^{-\sum_{k,l=1}^{N}\bar\xi_kS_{kl}\xi_l}=
\lim_{N\to\infty}\det S.$$
In our case the $N\times N$ matrix $S$ reads
\begin{eqnarray}
S= \left(\begin{array}{llll}
2 &0&...&b \\
-b&2&...&0\\
\vdots\\
0&...&-b&2\\
\end{array}
\right),
\end{eqnarray}
where $\quad b=1-\frac{\beta}{N}\overline\mu.$ The determinant is evaluated to yield
\begin{equation}
\lim_{N\to\infty}\det S=\lim_{N\to  \infty}\left[2^N+(1-\frac{\beta}{N}\overline\mu)^N\right].
\label{1.24}\end{equation}

If, in contrast, we absorb the extra factor produced by the measure
in the chemical potential term, we, instead, end up with
$$\lim_{N\to  \infty}\left[1+(1-\frac{\beta}{N}\overline\mu)^N\right]=1+e^{-\beta\overline\mu},$$
which yields the familiar result for non-interacting spinless fermions with the chemical potential
$$\mu\to\bar\mu=\mu+\frac{1}{\beta}\int_0^{\beta}ia^{(0)}d\tau.$$
The $2^N$ factor in Eq.(\ref{1.24}) is produced by the modification of the measure.
As discussed earlier in our analysis of the time discretized representation,
it cannot be absorbed in the chemical potential term.
Besides, integrating out the Fermi fields formally results in a divergent expression in the limit $N\to \infty.$
However, as we show below the remaining integral in (\ref{1.22}) over the bosonic fields $\bar z_k,z_k$ precisely
cancels such a divergence.

Explicitly, we get
$$tr e^{-\beta\mu X^{00}}=\lim_{N\to\infty}\int\prod_{k=1}^{N}\frac{d\bar z_kdz_k e^{ia_k}}
{2\pi i(1+\bar z_kz_k)^2}\det S=\lim_{N\to\infty}\int\prod_{k=1}^{N}\frac{d\bar z_kdz_k e^{ia_k}}
{2\pi i(1+\bar z_kz_k)^2}(2^N+e^{-\beta\mu}e^{-i\sum_ka_k})$$
$$=\lim_{N\to\infty}2^N\int\prod_{k=1}^{N}\frac{d\bar z_kdz_k}
{2\pi i(1+\bar z_kz_k)^2}e^{ia_k}
+e^{-\beta\mu},$$
since
\begin{equation}
\int \frac{d\bar zdz}{2\pi i(1+\bar zz)^2}=1.
\label{1.25}\end{equation}
Using the result
$$\int\prod_{k=1}^{N}\frac{d\bar z_kdz_k}
{2\pi i(1+\bar z_kz_k)^2}e^{ia_k}=\int\prod_1^N\frac{d\bar z_kdz_k}{2\pi i(1+\bar z_kz_k)^2}
\langle z_k|z_{k-1}\rangle\mid_{z_0=z_N}$$
\begin{equation}
=\int\prod_1^N\frac{d\bar z_kdz_k}{2\pi i(1+\bar z_kz_k)^2}\langle z_N|z_{N-1}\rangle\cdots
\langle z_1|z_N\rangle=2^{-N+1},
\label{2N}\end{equation}
where
\begin{equation}
\int \frac{(2s+1)d\bar zdz}{2\pi i(1+\bar zz)^2}|z\rangle_s\langle z|_s=
\int \frac{2d\bar zdz}{2\pi i(1+\bar zz)^2}|z\rangle_{s=1/2}\langle z|_{s=1/2}=I,
\label{f2}\end{equation}
finally gives
$$tr e^{-\beta\mu X^{00}}=2+e^{-\beta\mu},$$ as desired.
On the other hand, if we suppress the extra measure factor
we find instead the incorrect result,
$$tr e^{-\beta\mu X^{00}}=e^{-\beta\mu}.$$

At this point it is instructive to compare the action (\ref{1.20})
with its formal analog for
conventional spinless fermions coupled to $SU(2)$ spins
through the Berry's potential:
\begin{eqnarray}
S_{s/f}&=&\int_0^{\beta}ia^{(0)}(\tau)d\tau-\int_0^{\beta}\bar f
\left(\partial_{\tau}+\mu+ia^{(0)}\right)f d\tau.
\label{1.201}\end{eqnarray}
The standard path-integral representation of the pertinent partition function
reads
\begin{eqnarray}
Z_{s/f}&=&\int D\mu^{fermion}_{su(2)} e^{S_{s/f}},
\label{2.202}\end{eqnarray}
where
\begin{equation}
 D\mu^{fermion}_{su(2)}=
\prod_{\tau}\frac{2\,d\bar z(\tau)dz(\tau)}{2\pi i(1+\bar z z)^2} d\bar{f}(\tau)df(\tau)
\label{sfm}\end{equation}
is a standard measure in the CS space generated by the basis $|z\rangle_{s=1/2}\times |f\rangle.$
Here $|f\rangle$ stands for the normalized CS associated with the fermionic algebra,
\begin{equation}
|f\rangle=\frac{1}{\sqrt{1+\bar ff}}e^{f{\hat f}^{\dagger}}|0\rangle_F= \frac{1}{\sqrt{1+\bar ff}}
(|0\rangle_F + f|1\rangle_F),
\label{f}\end{equation}
where  $\{\hat f,\hat f^{\dagger}\}=1.$ Within Shankar's approach, $\hat f^{\dagger}$ represents a hole creation
operator.
Accordingly, the one-particle state $|1\rangle_F$ corresponds to a hole excitation, whereas $|0\rangle_F$ stands
for a hole-vacuum
state. However, this vacuum state is unphysical, since it is not present in  the $3D$ on-site reduced  Hilbert space
for
strongly correlated electrons which is spanned by the spin-up, spin-down and hole (vacancy) state vectors.

The integral (\ref{2.202}) can be calculated on the time lattice along the lines depicted earlier on and yields
the divergent result,
\begin{equation}
Z_{s/f}=\lim_{N \to\infty}2(1+2^{N-1}e^{-\beta\mu}).
\label{diverg}\end{equation}
This clearly indicates that action (\ref{1.201}) cannot  represent a physical system. The
Berry's potential in the last term of Eq. (\ref{1.201})
takes into account the fact that the $su(2)$ spins and spinless fermions are correlated within Shankar's theory
due to the NDO constraint, even
in the absence of a direct interaction between them.
However,
that correlation is not taken into a full account there.
The projection of the spin-fermion Hamiltonian onto the physically
constrained Hilbert space
manifests itself both in the appearance of the Berry's phase in the fermionic action as well as
in the modification of the measure in the path integral (\ref{1.14}) for the partition function.

It is also important to realize that
the spin part of the measure in Eq.(\ref{sfm}) contains the extra factor of $2=(2s+1)_{s=1/2}$
in the numerator
compared to the spin part of the on-site version of Eq.(\ref{1.15}).
It comes
from the normalized $SU(2)$ spin measure as  given
by the resolution of the identity in the spin space (\ref{f2}).
The fermionic part of the measure (\ref{sfm}) is standard and comes from
the resolution of the identity in the fermionic Hilbert space,
$$\int\,d\mu_{fermion}\,|f\rangle\langle f|=
\int\,d\bar{f}df |f\rangle\langle f|=1.$$
This factorization of the path-integral measure (\ref{sfm}) into pure spin and fermionic parts reflects
the fact that the Hilbert space of the whole system is represented by a direct product of the spin
and fermion subspaces.
In contrast, the measure (\ref{1.15}) comes from the resolution of the unity in the whole superspace
$CP^{1|1}$ as given by Eq.(\ref{1.8}). It
cannot be factorized into the spin-fermion parts and the resulting effective action
represents a unique composed object. For example, the integration over
just the fermionic amplitudes in (\ref{1.22}) diverges.

Finally, let us evaluate the Green's function
\begin{equation}
G^{(0)}(\tau_q-\tau_r)=Z_0^{-1}\int D\mu (z,\xi )\xi(\tau_q)\bar\xi(\tau_r)
e^{{\cal S}_0(z, \xi)}, \label{1.26}\end{equation}
where
\begin{equation}
D\mu (z,\xi )=\prod_{\tau}\frac{d\bar z(\tau)dz(\tau)d\bar\xi(\tau)d\xi(\tau)}{2\pi i(1+|z|^2)^2}\,
(1-\bar\xi\xi), \quad z(0)=z(\beta),\, \xi(0)= -\xi(\beta)
\label{1.27} \end{equation} and
\begin{eqnarray}
{\cal S}_0=\int_0^{\beta}ia^{(0)}(\tau)d\tau-\int_0^{\beta}\bar\xi
\left(\partial_{\tau}+\mu+ia^{(0)}\right)\xi d\tau.
\label{1.28}\end{eqnarray}
Making the change of variables,
$\xi(\tau)\to \xi(\tau)e^{-i\phi(\tau)},$ where $\phi(\tau)$ is given by (\ref{phase}),
this can be brought into the form
\begin{eqnarray}
G^{(0)}(\tau_q-\tau_r)
= Z_0^{-1}\int D\mu (z,\xi )\xi(\tau_q)\bar\xi(\tau_r)e^{-i\phi(\tau_q)+i\phi(\tau_r)}
e^{{\cal S}_0(z, \xi)}, \label{1.29}\end{eqnarray}
where $\tau_{\alpha}=\frac{\beta}{N}\alpha$, $\alpha$ is integer, and ${\cal S}_0$ is given by (\ref{1.21}).
This expression is a continuum limit of the time-discretized representation
\begin{eqnarray}
G^{(0)}(\tau_q-\tau_r)
= Z_0^{-1}
\lim_{N\to\infty}\int\prod_{k=1}^{N}\frac{d\bar z_kdz_k e^{ia_k -i\phi(\tau_q)+i\phi(\tau_r)}}
{2\pi i(1+\bar z_kz_k)^2}(\det S) S^{-1}_{qr},
\label{1.31}\end{eqnarray}
where the inverse of $S$ is
\begin{eqnarray}
S^{-1}_{qr}= (\det S)^{-1}\left(\begin{array}{llll}
2^{N-1} &-b^{N-1}&...&-b2^{N-2} \\
2^{N-2}b&2^{N-1}&...&-b^22^{N-1}\\
\vdots\\
b^{N-1}&2b^{N-2}&...&2^{N-1}\\
\end{array}
\right), \quad \det S=2^N+b^N.
\end{eqnarray}
Hence, for $q>r$
\begin{equation}
S^{-1}_{qr}=\frac{2^{N-1-(q-r)}b^{q-r}}{2^N+b^N},
\label{1.32}\end{equation}
and this gives
\begin{eqnarray}
G^{(0)}(\tau_q-\tau_r)
&=& (2+e^{-\beta\mu})^{-1}
\lim_{N\to\infty}\int\prod_{k=1}^{N}\frac{d\bar z_kdz_k }
{2\pi i(1+\bar z_kz_k)^2}2^{N-1-(q-r)}e^{-\mu(\tau_q-\tau_r)}\nonumber\\
&\times&\prod_{k=1}^N\langle z_k|z_{k-1}\rangle
(\prod_{k=r}^q\langle z_k|z_{k-1}\rangle)^{-1}.
\label{1.33}\end{eqnarray}
Since
$$\int\prod_{k=1}^{N}\frac{d\bar z_kdz_k }
{2\pi i(1+\bar z_kz_k)^2}\prod_{k=1}^N\langle z_k|z_{k-1}\rangle
(\prod_{k=r}^q\langle z_k|z_{k-1}\rangle)^{-1}=2^{-N+(q-r)+2},$$
we finally get
\begin{equation}
G^{(0)}(\tau_q-\tau_r)\mid_{q>r}=e^{-\mu(\tau_q-\tau_r)}(1-\delta),
\label{1.34}\end{equation}
where the vacancy occupation probability $\delta$ is determined by
$$\delta=\langle X^{00}\rangle=\frac{1}{2e^{\beta\mu}+1}.$$
Similarly, for $q<r$, we find
\begin{equation}
G^{(0)}(\tau_q-\tau_r)\mid_{q<r}=-e^{-\mu(\tau_q-\tau_r)}\delta.
\label{1.35}\end{equation}
Combining these results, we finally get
\begin{equation}
G^{(0)}(\tau-\tau')=e^{-\mu(\tau-\tau')}\left[\theta(\tau-\tau'-\eta)(1-\delta)-\theta(\tau'-\tau+\eta)\delta\right],
\label{1.36}\end{equation}
where the infinitesimal parameter $\eta=0^{+}$ is a reminder  that
the variables $\bar\xi(\tau')$ and $\xi(\tau)$ in the path integral (\ref{1.14}) are associated with variables
displaced by one time step, $\bar\xi(\epsilon k)$ and $\xi(\epsilon(k-1))$, respectively.
At equal times we find
\begin{equation}
-G^{(0)}(0^-)=
\langle \bar\xi(\tau)\xi(\tau)\rangle=\langle X^{00}\rangle=\delta,
\label{1.37}\end{equation}
as it should. In contrast, for the conventional spinless
fermions governed by the Hamiltonian $H=\mu \hat f^{\dagger}\hat f$ one obtains
Eq.(\ref{1.36}) with $\delta$ replaced by $ n_F$.

Because of the extra factor $\prod_{\tau}(1-\bar\xi(\tau)\xi(\tau))$ in the measure
(\ref{1.15}), the fields $\bar\xi,\xi$ separately
do not have a direct physical meaning. They don't represent spinless fermions, for instance.
However, the bilinear combination $\bar\xi(\tau)\xi(\tau')$
represents, at equal times, a physical observable, the covariant symbol of the vacancy number operator $X^{00}$.
We see that
the spin degrees of freedom are nontrivially involved in the evaluation of a purely fermionic vacancy propagator
(\ref{1.36}).
Notice, however, that this spin-charge correlation due to the NDO constraint does not prevent the spin-charge
separation observed in the $1D$ Hubbard model.
The physical elementary excitations that represent separately the spin and charge degrees of freedom in the $1D$
Hubbard model are not simply the $z$ and $\xi$ field excitations, but
are, instead, nonlocal string objects composed, simultaneously, of both types of  bare elementary excitations \cite{weng}.

Finally, let us say a few words concerning the meaning of the continuum path-integral representation (\ref{1.14}).
Strictly speaking, Eq.(\ref{1.14}) is only symbolic: the measure in that path integral cannot
be defined in a mathematically rigorous way. This observation has a deep physical meaning. If a
continuum path integral could be defined as a bona-fide integral with respect to a genuine measure insensitive
to discrete approximations, it
would immediately provide a one-to-one correspondence between classical and quantum physics
thereby making the latter
an unnecessary ingredient. However, classical dynamics is known to give rise to different quantum
theories if one follows different quantization schemes. A specific quantization scheme is encoded into
a discrete representation of the continuum path integral that should be taken as its true definition.
In our case the basic discrete representation  is given by our Eq.(\ref{1.12}).
Since
\begin{equation}
\xi(\tau+\eta)=\xi(\tau)+\frac{d\xi(\tau)}{d\tau}\eta+{\cal O}(\eta^2),\quad \eta\to 0,
\label{sf}\end{equation}
it might seem that one could safely ignore that shift in the low-energy limit,
$\frac{d\xi}{d\tau}\propto \omega \to 0,$ where $\omega$ stands for a characteristic frequency of
the fermionic degrees of freedom
\cite{shankar2}. However, this is not the case: the path-integral variables are not in general smooth functions.
There may exist different discrete approximations to one and the same
continuum action that result in completely different low-energy dynamics.
Some explicit examples can be found in \cite{negele}.
Even in cases where the continuum limit of a path integral does make sense, i.e. in a semiclassical
or perturbation theory,
the relative shift of the arguments of the path-integral variables still cannot
be ignored.
The discontinuity of the correlators such as $<\xi(\tau)\bar\xi(\tau')>$,
at equal
time arguments, should be dealt with according to rules following from the defining discrete approximation.
As soon as the discrete approximation is fixed, no "operator ordering problem" shows up. In our case the order
is fixed from the very beginning and manifests itself in the Hamiltonian function defined by Eq.(\ref{mar}).

\section{1D example}

The $U=\infty$ Hubbard model (\ref{1.1}) is known to be exactly solvable in $1D$.
We show below  that the exact
ground-state energy can be recovered within
the path-integral representation (\ref{1.14}-\ref{1.16}).
To start with,
the ground state of the $1D$  $U=\infty$  Hubbard model is known to be degenerate with respect to
spin. To calculate the path integral we can therefore choose any spin configuration, e.g., the ferromagnetic (FM)
one. We thus  put $z_i=z_j$ in Eq. (\ref{1.17}), which  yields
\begin{equation}
Z_{U=\infty}=\int D\mu (z,\xi )\
e^{S(z, \xi)}, \label{e1}
\end{equation}
where the measure is given by Eq.(\ref{1.15}) and
the action reads
\begin{eqnarray}
S=\int_0^{\beta}tr\,\, ia^{(0)}d\tau-\int_0^{\beta}\bar\chi
\left(\partial_{\tau}+ia^{(0)}+T\right)\chi d\tau.
\label{e2}\end{eqnarray}
Here $T$ and $a^{(0)}$ are  the $N_s\times N_s$ matrices,
\begin{equation}
(a^{(0)})_{ij}=a^{(0)}_i\delta_{ij}, \quad (T)_{ij}=t_{ij}-\mu\delta_{ij},
\label{T}\end{equation}
with $i,j=1,2,..,N_s$ numbering the lattice sites. The vector $\chi=(\xi_1,\xi_2,..,\xi_{N_s})^t$, and the trace is taken
over the lattice site indices.
On the time lattice, the integral over the $\chi$ fields can be carried out to yield
$$Z_{U=\infty}=\lim_{N\to\infty}\int\prod_{k=1}^{N}\prod_j^{N_s}\frac{d\bar z_k(j)dz_k(j)\det\, e^{ia_k}}
{2\pi i(1+\bar z_k(j)z_k(j))^2}\det (2^N+e^{-\beta T -i\sum_ka_k})$$
\begin{equation}
=\lim_{N\to\infty}\int\prod_{k=1}^{N}\prod_j^{N_s}\frac{d\bar z_k(j)dz_k(j)}
{2\pi i(1+\bar z_k(j)z_k(j))^2}
\det\,(2^N e^{i\sum_ka_k}+e^{-\beta T}), \quad z_k(j)\equiv z_j(\epsilon k).
\label{1d}\end{equation}

Since the matrix $||2^N e^{i\sum_ka_k}+e^{\beta\mu }||$ at $t_{ij}=0$ is diagonal,
Eq.(\ref{1d}) reduces to
$$Z_{U=\infty}(t=0)=(2+e^{\beta\mu})^{N_s}=e^{\beta\mu N_s}(1+o(1)), \, \beta\to\infty,\, \mu>0,$$
where $N_s$ is fixed. This result is a direct consequence of Eq. (\ref{2N}).
Accordingly, at $t_{ij}\neq 0$ the negative eigenvalues of the matrix $T$ determine the asymptotic
behaviour of the partition function.
Consequently,
$$Z_{U=\infty}=\prod_{p:\,T_p<0}e^{-\beta T_p}(1+o(1)), \quad  \beta\to\infty,$$
where $T_p=-t_p-\mu, \,\, t_p=2t\cos p,\,\, p \in BZ.$
The ground-state energy then
becomes
\begin{equation}
E_{gr}/N_{s}=-\frac{2t}{\pi}\sin(\pi\delta),\quad \delta=\frac{1}{\pi}\arccos(\frac{-\mu}{2t}), \, t\ge 0,
\label{e3}\end{equation}
which coincides with the exact  $1D$ result for
the $U=\infty$ Hubbard model \cite{ogata}.

\section{Doped antiferromagnet}

Let us now turn our attention to a derivation of the low-energy effective action of
a doped AF, starting right from the microscopic $t-J$ model,
\begin{equation} H_{t-J}=H_t+H_J=-t\sum_{ij\sigma} \tilde{c}_{i\sigma}^{\dagger}
\tilde{c}_{j\sigma}+ J\sum_{ij} \vec Q_i \cdot \vec Q_j,
\label{2.1}
\end{equation}
where $\vec Q_i$ stands for the local electron spin operators given by Eq.(\ref{qx})
and $J\ge 0$ describes the nn exchange interaction.
The parameter $J\sim {\cal O}(1/U)$ and the bare constants are chosen such that
$t\gg J.$

Because of the NDO constraint there are no charge fluctuations
at half filling $(\delta=0)$, and precisely in this limit, the $t-J$ model reduces to a  Heisenberg AF model
\begin{equation}
H_{t-J}\to H^{\delta=0}_{J}= J\sum_{ij} \vec S_i \cdot \vec S_j,
\label{2.2}
\end{equation}
with no restriction on $J$, the sole energy scale, apart from its positive sign.
In the low-energy long-wavelength limit $H_J$ gives rise to the action of the
nonlinear sigma-model.

Proceeding as discussed in the preceding section we arrive at the
representation of the $t-J$ partition function,
\begin{equation}
Z_{t-J}=\int D\mu (z,\xi )\
e^{\int_0^{\beta}L_{t-J}(z, \xi)d\tau}, \label{2.3}
\end{equation}
where the measure factor $D\mu (z,\xi )$ is given by Eq.(\ref{1.15}).
The Lagrangian now reads
\begin{eqnarray}
L_{t-J}&=&\sum_iia^{(0)}_i(\tau)-\sum_i\bar\xi_i
\left(\partial_{\tau}+\mu+ia^{(0)}_i\right)\xi_i-
H^{cov},
\label{2.4}\end{eqnarray}
where
\begin{eqnarray}
H^{cov}=
t\sum_{i\ne j}\overline{\xi }_{j}\xi _{i}\langle
z_{i}|z_{j}\rangle +J\sum_{i\ne j}\vec S^{cov}_i\vec S^{cov}_j(1-\bar\xi_i\xi_i)(1-\bar\xi_j\xi_j).
\label{2.5}
\end{eqnarray}
The summation in (\ref{2.5}) is extended over nn and nnn sites.  As a result we find
$$\langle z_{i}|z_{j}\rangle=\langle z(\vec r_i)|z(\vec r_i+\delta\vec r)\rangle=1+\langle z(\vec r_i)|\frac{d}{d\vec r_i}
|z(\vec r_i)\rangle\delta\vec r +{\cal O}(\delta\vec r^2)= 1+\langle z(\vec r_i)|
\frac{\partial z_i}{\partial \vec r_i}\frac{\partial}{\partial z_i}
+\frac{\partial \bar z_i}{\partial \vec r_i}
\frac{\partial}{\partial\bar z_i}|z(\vec r_i)\rangle\delta \vec r +{\cal O}(\delta\vec r^2)$$
$$=1-i\vec a_i\delta\vec r+{\cal O}(\delta\vec r^2)=\exp(-i\vec a_i\delta\vec r)+{\cal O}(\delta\vec r^2).$$
where $\vec r_j=\vec r_i+\delta\vec r.$
Vector $\vec a$ represents the spatial components of the pull-back of the Berry's connection one-form,
$\varphi^*A=a,$ with $\varphi:(\tau,\vec r) \to z(\tau,\vec r)$ and
$$A=i\langle z|d|z\rangle.$$ Here $d$ stands for the exterior derivative,
$$d=dz\frac{\partial}{\partial z}+d\bar z\frac{\partial}{\partial \bar z}.$$

The effective hole action becomes
\begin{eqnarray}
S_t&=&\sum_{\vec r_i}\int_0^{\beta}ia_i^{(0)}(\tau)d\tau-\sum_{\vec r_i}\int_0^{\beta}\bar\xi_{\vec r_i}
\left(\partial_{\tau}+\mu+ia^{(0)}_i\right)\xi_{\vec r_i}d\tau\nonumber\\&-&
t\int_0^{\beta}\sum_{\vec r_i, \delta\vec r}\overline{\xi }_{\vec r_i+\delta\vec r}\xi _{\vec r_i}
e^{-i\vec a_i\delta\vec r}d\tau +{\cal O}(\delta\vec r^2).
\label{2.6}\end{eqnarray}
Here
$a_i^{\nu}(\tau):=a^{\nu}(\vec r_i,\tau),\,\nu=0, x,y,$ with $i\vec a_i= -\langle z_i|
\frac{\partial}{\partial \vec r_i}|z_i\rangle.$ Note also that $z_i:=z(\vec r_i)$ and
$\vec r_j=\vec r_i+\delta\vec r$ where $\delta\vec r\propto a$ with $a$ being the lattice spacing.

The whole action should be complemented by the $J$-term. Right at half filling the only surviving
term
gives rise to the nonlinear sigma-model action to describe the quantum AF. The perfect Neel ordering prevents
interlattice hole hopping. This follows directly from Eq. (\ref{2.5}). Namely, the AF long-range order,
$\vec S_i=-\vec S_j=\vec S $, implies
\begin{equation}
z_i=z,\, z_j=-1/\bar z,\quad i\in A, j\in B.
\label{2.7}\end{equation}
In view of (\ref{cs}), this
results in $\langle z_{i\in A}|z_{j\in B}\rangle=
\langle z|-1/\bar z\rangle=0,$ and there is no hopping between the $A$ and $B$ sublattices. Only
intralattice nnn hopping of vacancies is possible in this case.
One can also check that
\begin{equation}
\langle z|d|z\rangle=-\langle -1/\bar z|d|-1/\bar z\rangle,
\label{2.8}\end{equation}
which means that $a_{i\in A}^{\nu}=-a_{i\in B}^{\nu}.$

Suppose now that we lightly  dope
the AF with holes, with the dominant energy scale in the problem continuing to be the
exchange coupling $J$.
We assume that the AF ordered lattice spins are slightly perturbed by a small amount of vacancies.
In this case for a small enough hole concentration
$\delta$, one gets
$$z_i=-1/\bar z_j +{\cal O}(\delta),$$
where $i,j$ are the nn sites. This in turn implies
$$\langle z_i|z_j\rangle={\cal O}(\delta),\quad  \delta\to 0.$$
Therefore , very close to half filling, the interlattice hopping effectively results in the renormalization
of the hopping amplitude $t\to \delta t$ and, as a result, this term can be discarded.

Summing this all up, the partition function of the $t-J$ model close to half filling reads
\begin{equation}
Z_{t-J}=\int D\mu (z,\xi )\
e^{\int_0^{\beta}L_{t-J}(z, \xi)d\tau}. \label{2.10}
\end{equation}
Here
\begin{eqnarray}
L_{t-J}&=&\sum_{\vec r_i\in A}ia_i^{(0)}(\tau)-\sum_{\vec r_i\in A}\bar\xi_{\vec r_i}
\left(\partial_{\tau}+\mu+ia^{(0)}_i\right)\xi_{\vec r_i}\nonumber\\&-&
t\sum_{\vec r_i, \delta\vec r\in A}\overline{\xi }_{\vec r_i+\delta\vec r}\xi _{\vec r_i}
e^{-i\vec a_i\delta\vec r} +(A\to B, a_{i}\to -a_i)\nonumber\\
&-& \tilde J\sum_{\vec r_i,\delta\vec r}\vec S^{cov}_{\vec r_i}\vec S^{cov}_{\vec r_i+\delta\vec r},
\quad \tilde J=J(1-\delta)^2,
\label{2.9}\end{eqnarray}
and the measure is given by Eq.(\ref{1.15}),
$$D\mu (z,\xi )=\prod_{i,\tau}\frac{d\bar z_i(\tau)dz_i(\tau)}{2\pi i(1+|z_i|^2)^2}\,
d\bar\xi_i(\tau)d\xi_i(\tau)(1-\bar\xi_i\xi_i).$$

Within Shankar's approach, the underdoped AF is described by the partition function
\begin{equation}
Z^{Shankar}=\int D\mu_{su(2)}^{fermion}(z,f)\
e^{\int_0^{\beta}L^{Shankar}(z, f)d\tau}. \label{2.12}
\end{equation}
Here
\begin{eqnarray}
L^{Shankar}&=&\sum_{\vec r_i\in A}ia_i^{(0)}(\tau)-\sum_{\vec r_i\in A}\bar f_{\vec r_i}
\left(\partial_{\tau}+\mu+ia^{(0)}_i\right)f_{\vec r_i}\nonumber\\&-&
t\sum_{\vec r_i, \delta\vec r\in A}\bar f_{\vec r_i+\delta\vec r} f_{\vec r_i}
e^{-i\vec a_i\delta\vec r}+(A\to B, a_{i}\to -a_i)\nonumber\\
&-& \tilde J\sum_{\vec r_i,\delta\vec r}\vec S_{\vec r_i}^{cov}\vec S_{\vec r_i+\delta\vec r}^{cov},
\label{2.122}\end{eqnarray}
and the unconstrained spin-fermion measure reduces to
\begin{equation}
 D\mu_{su(2)}^{fermion}=
\prod_{i\tau}\frac{2\,d\bar z_i(\tau)dz_i(\tau)}{2\pi i(1+\bar z_i z_i)^2} d\bar{f}_i(\tau)df_i(\tau).
\label{sfm1}\end{equation}
Here the amplitudes denoted by $f_i$ describe the conventional spinless fermions, whereas the $z_i$
fields correspond to the $su(2)$ spins.
This is Shankar's result for the underdoped AF (with $s=1/2$).

Formally, the Lagragians (\ref{2.9}) and (\ref{2.122}) appear to be identical.
However,
due to the different measures, the corresponding partition functions
are completely different from each other.
The fermionic extra term in our measure (\ref{1.15}) cannot be ignored.
Physically, it reflects a rearrangement of the underlying Hilbert space induced by the NDO constraint.
This clearly cannot be treated perturbatively.
In a true spin-fermion Hamiltonian,
the on-site Hilbert space is represented by a $4D$ direct product of the $2D$ spin and the $2D$ spinless fermion subspaces.
However, the NDO constraint  reduces it to a $3D$ on-site Hilbert space composed of three state vectors: the spin-up,
spin-down and the vacancy states. As shown above, it is precisely the extra measure term that takes explicit care of
that distinction.

As an example, consider  partition functions (\ref{2.10}) and (\ref{2.12})
in the limiting case $t=J=0$:
\begin{equation}
L_{t-J}(t=J=0)=\sum_{\vec r_i}ia_i^{(0)}(\tau)-\sum_{\vec r_i}\bar\xi_{\vec r_i}
\left(\partial_{\tau}+\mu+ia^{(0)}_i\right)\xi_{\vec r_i},
\label{e1}\end{equation}
and
\begin{equation}
L^{Shankar}(t=J=0)=\sum_{\vec r_i}ia_i^{(0)}(\tau)-\sum_{\vec r_i}\bar f_{\vec r_i}
\left(\partial_{\tau}+\mu+ia^{(0)}_i\right)f_{\vec r_i}.
\label{e2}\end{equation}
In view of Eqs.(\ref{1.22})
$$Z_{t-J}(t=J=0)=(2+e^{-\beta\mu})^{N_s},$$ as it should.
On the other hand, Eq.(\ref{diverg}) tells us that $Z^{Shankar}(t=J=0)$ diverges,
which is inappropriate for a system with a finite
number of degrees of freedom. If one, however, drops the $a^{(0)}_i$ potential in the fermionic part of Eq.(\ref{e2}),
one arrives at the conventional partition function
$$2^{N_s}(1+e^{-\beta\mu})^{N_s}$$
which describes entirely independent uncorrelated spin and fermionic degrees of freedom.

Only right at half filling, $\delta=0$, our representation (\ref{2.10}) and Shankar's theory
become identical.
This can be seen as follows.
Since there are no holes in this limit, the fermionic amplitudes $(f, \bar f)$ describing their propagation
throughout the lattice vanish identically.
The partition function then becomes
\begin{equation}
Z^{Shankar}_{\delta=0}=\int D\mu_{su(2)}\,
e^{\int_0^{\beta}L^{Shankar}(\delta=0)d\tau}, \label{3.1}
\end{equation}
where
\begin{eqnarray}
L^{Shankar}(\delta=0)=\sum_i ia^{(0)}_i(\tau)-H^{cov},\quad H^{cov}= J\sum_{i\neq j}
\vec S^{cov}_{i}\vec S^{cov}_{j},
\label{3.2}\end{eqnarray}
and the normalized spin measure is given by
\begin{equation}
D\mu_{su(2)}=\prod_{i,\tau}\frac{2d\bar z_i(\tau)dz_i(\tau)}{2\pi i(1+|z_i|^2)^2}.
\label{3.3}\end{equation}

On the other hand, at $t=0$, the projected $t-J$ Hamiltonian reads
\begin{equation}
H_J= J\sum_{ij} \vec Q_i \cdot \vec Q_j +\mu\sum_i X^{00}_i.
\label{3.4}
\end{equation}
The half filling limit in this representation can be enforced
by sending the chemical potential $\mu$ to $+\infty$. This has an immediate
effect in allowing only the zero eigenvalues of the local
vacancy number
operator $X^{00}_i$ to survive.
In this way the partition function becomes
\begin{equation}
Z_{J}=\int D\mu (z,\xi )\
e^{\int_0^{\beta}L_{J}d\tau}, \label{3.5}
\end{equation}
with
\begin{eqnarray}
L_{J}&=&\sum_{i}ia_i^{(0)}(\tau)-\sum_{_i}\bar\xi_{i}
\left(\partial_{\tau}+\mu+ia^{(0)}_i\right)\xi_{i}\nonumber\\&-&
J\sum_{i\neq j}\vec S^{cov}_{i}\vec S^{cov}_{j},\quad \mu\to +\infty.
\label{3.6}\end{eqnarray}
The measure $D\mu (z,\xi )$ is given by Eq.(\ref{1.15}). Note that it is normalized by the condition (\ref{1.8}),
so that
the factor of $2=(2s+1)_{s=1/2}$ is missing in the numerator of its spin part.

Notice that we cannot simply put $\xi=\bar \xi=0$ in Eq.(\ref{3.6}). This is  because those fields have no physical
meaning and do not
represent holes directly as opposed to the amplitudes $f, \bar f$. We must instead integrate them out in the limit
$\mu\to +\infty$.
We do that on the time lattice, which gives
\begin{equation}
Z_J=\lim_{N\to\infty}2^{NN_s}\int\prod_{k=1}^{N}\prod_j^{N_s}\frac{d\bar z_k(j)dz_k(j)\det\, e^{ia_k}}
{2\pi i(1+\bar z_k(j)z_k(j))^2}\exp{[-\epsilon\sum_k H_{cov}(k,k-1)]}.
\label{3.7}\end{equation}
Here $$ H_{cov}(k,k-1)=
 J\sum_{i\neq j}\vec S_{i}^{cov}(\bar z_k,z_{k-1})\vec S_{j}^{cov}(\bar z_k,z_{k-1}).$$ Note that the factor
of $2^{NN_s}$ comes precisely from the
fermionic extra factor in the measure (\ref{1.15}).
This factor can be absorbed back into the spin measure to turn it into a conventional normalized spin measure
(\ref{3.3}). As a result, the integral over
the spin fields $z, \bar z$ in (\ref{3.7}) becomes identical to that given by Eq. (\ref{3.1}).
The distinction between the representations ({\ref{2.10}) and (\ref{2.12}) disappears  at half filling since
the corresponding on-site spin Hilbert spaces become identical.

Finally, the Berry's gauge potential $a^{\nu}=(\varphi^*A)^{\nu}$ appears in the path-integral
action (\ref{2.9}) as an external gauge field and has no dynamical role.
The gauge-theory approaches to treat the
$t-J$ model basically fall into two categories.
The first one emerges from the
slave-particle representations of the constrained electron operators.
The idea of that approach is to re-express these  constrained operators in terms of
the standard boson/fermion bilinears. This is equivalent to the so-called oscillatory representations
of $su(2|1)$ superalgebra.
For example, the
physical electron operator can be represented by a product of a spinful boson and slave spinless fermion
$$\tilde c_{i\sigma}=f_ib^{\dagger}_{i\sigma},$$ with standard commutation/anticommutations rules.
However, this representation clearly increases the on-site number of degrees
of freedom by a factor of $2$. The emergent $U(1)$ local gauge field, $f_i,b_{i\sigma}\to e^{i\theta_i} f_i, e^{i\theta_i}
b_{i\sigma}$,
takes care of one redundant degrees of freedom (by fixing a gauge), while
the NDO constraint
$$f^{\dagger}_if_i+\sum_{\sigma}b^{\dagger}_{i\sigma}b_{i\sigma}=1$$
takes care of the other. However, the elementary excitations of the slave-particle
fields do not represent physical excitations since they are gauge dependent.
This compact dynamical $U(1)$ gauge field is generated by the NDO constraint.
Within our approach the constraint is resolved explicitly and there is no
need for an imposing the $U(1)$ gauge field theory. In fact, an explicit resolution of the NDO constraint within the slave-fermion theory
results exactly in the $su(2|1)$ path-integral representation \cite{k2}.

The gauge potential $a^{\nu}_i=(\varphi^*A_i)^{\nu}$ which appears in Eq.(\ref{2.9}) has a different nature
altogether.
It is driven by the fluctuations of the spin background
present in any $su(2)$ path-integral action regardless of the chosen coordinates and it
takes care of the $U(1)$ redundancy of the spin quantum state $|z\rangle$.
From the geometric viewpoint, this can be stated as follows \cite{stone}.
The $su(2)$ CS's can be considered as sections of the
principle $U(1)$ bundle
$P(CP^1,U(1))$ frequently referred to as a magnetic monopole bundle.
The base space of this bundle, $M= CP^1$, appears as a classical phase space of the spin, whereas
its covariantly constant sections $|z\rangle$ form a quantum
Hilbert space for the spin, with
$$ \nabla |z\rangle=0,\quad  \nabla:=d+iA.$$
Since the manifold $CP^1\simeq S^2$ is topologically nontrivial, the monopole bundle is nontrivial
as well. Notice that the one-form $A$ does not exist globally, and
any two locally defined gauge potentials are related by a $U(1)$ gauge transformation.
Fixing the spin CS by  Eq. (\ref{1.2a}) amounts to fixing a local section of the bundle.

Under the global  canonical $SU(2)$ transformations acting in
the base space we get
\begin{eqnarray}
z_i \rightarrow z_i^g=
\frac{uz_i+v}{-\overline{v}z_i+\overline{u}},\quad
g= \left(\begin{array}{ll}
u & v \\
-\overline{v} & \overline{u}%
\end{array}
\right) \in \mathrm{SU(2}).
\label{2.177}\end{eqnarray}
When lifted to the bundle (quantum) space, this gives
\begin{equation}
g\to U_g:\quad U_g|z_i\rangle=e^{-i\zeta_i}|z_i^g\rangle, \quad \zeta_i=-i\log \sqrt{\frac{-v\overline{z}_i+u}
{-\overline{v}z_i+\overline{u}}}.
\label{2.77}\end{equation}
The Berry's connection transforms according to
\begin{equation}
A_i\to A_i+d\zeta_i.
\label{2.17}\end{equation}
and the Fermi field becomes
\begin{eqnarray}
\xi_i\rightarrow
e^{i\zeta_i }\xi_i,
\label{2.18}
\end{eqnarray}
leaving the whole action in Eq.(\ref{2.10}) globally $SU(2)$ invariant.

\section{Discussion}

In this section we discuss some physical consequences
that follow from representations (\ref{2.10}) and (\ref{2.12}).
Let us start with Shankar's theory (\ref{2.12}).
This theory naturally allows for the hopping of the conventional fermions throughout the lattice
in the absence of the local spin degrees of freedom. The corresponding partition function
follows from the representation (\ref{2.12}) if one discards the spin degrees of freedom.
In the momentum representation, it reads
\begin{equation}
Z^{Shankar}_{\vec S_i=0}=\int \prod_{\vec p,\tau}D\bar f_{\vec p}(\tau)Df_{\vec p}(\tau)
\exp{\left(-\sum_{\vec p}\int_0^{\beta}\bar f_{\vec p}(\partial_{\tau}-\epsilon_{\vec p})f_{\vec p}d\tau\right)}.
\label{4.01}
\end{equation}
Here $\epsilon_{\vec p}=t_{\vec p}-\mu,$ and $t_{\vec p}=2t\sum_{\vec a}\cos (\vec p\,\vec a)$, where $\vec a$
is a lattice vector and $\vec p \in BZ$. This path integral can be easily computed to
yield a partition function for the conventional spinless fermions,
$$Z^{Shankar}_{\vec S_i=0}=\prod_{\vec p}\left(1+e^{\beta\epsilon_{\vec p}}\right).$$

Having this in mind,
the low-energy long-wavelength limit can be taken explicitly
to reduce Shankar's action (\ref{2.122}) in $1D$ to that of Dirac fermions coupled to the nonlinear sigma model
via a compact $U(1)$ gauge field \cite{shankar}. Following the usual procedure for $1D$ systems
to take into account the low-energy fermionic
excitations, we linearize the theory near
the Fermi points $\pm k_F$. The spinless Fermi
amplitudes $\psi=f/\sqrt{a}$ can be written as follows:
\begin{equation}
\psi(n)=e^{ik_Fn}\psi_1(n)+e^{-ik_Fn}\psi_2(n).
\label{4.1}
\end{equation}
Here index $n$ stands for the lattice sites. The resulting action in the continuum
limit reads
\begin{eqnarray}
Z^{1D}_{Shankar}&=&\int D\bar\psi D\psi D\mu_{su(2)}(\bar z,z)e^{S_F+S_{\theta}},\nonumber\\
S_F&=&\int[\bar\psi_A(-\hat{\partial} -i\hat a)\psi_A+ \bar\psi_B(-\hat\partial+i\hat a)\psi_B]dxd\tau,\nonumber\\
S_{\theta}&=& -\frac{1}{2}\int dxd\tau(c\partial_x\bar z\partial_x z+c^{-1}\dot{\bar z}\dot z)+i\theta W,\quad c=Ja,
\label{4.2}\end{eqnarray}
where $S_{\theta}$ is the sigma model action including the topological $\theta$-term
and $\hat a = a^{\nu}\gamma_{\nu}.$ The chemical potential $\mu$ is incorporated in the theory
through the relation $k_F=\arccos(-\mu/2t)$.
The Euclidean $2\times 2$ gamma matrices
can be taken in the form
$\gamma_0=\sigma_y,\, \gamma_1=\sigma_x$ so that $\gamma_5=i\gamma_0\gamma_1=\sigma_z.$
Notice also that $\psi =(\psi_1, \psi_2)^t, \quad \bar\psi=\psi^{\dagger}\gamma_0.$

An interesting observation concerning the representation (\ref{4.2}) is that the
dependence on the parameter $\theta$ actually drops out from the theory \cite{shankar}.
This can be seen as follows.
The action to describe the massless fermions coupled with the spin background fields $z,\bar z$,
$$ S_F=\int[\bar\psi(-\hat{\partial} \pm i\hat a)\psi]dxd\tau,$$ is invariant under a chiral $U(1)$
transformation,
\begin{equation}
\psi\to e^{i\gamma_5\phi}\psi,\quad \bar\psi\to \bar\psi e^{i\gamma_5\phi},
\label{4.3}\end{equation}
with $\phi$ being a parameter. However, this is no longer the case for the partition function
\begin{equation}
Z_{F}=\int D\bar\psi D\psi e^{S_F}.
\label{4.31}\end{equation}
The fields $z(x,\tau),\bar z(x,\tau)$ map a compactified $2D$
plane $(x,\tau)$ homeomorphic to a two sphere $S^2$ onto a spin phase space which is also a two sphere,
$S^2\to S^2$ It is known that such maps can be classified by the integers $W$ which define the number of times
the the ``space-time'' sphere covers the ``spin'' sphere.
Explicitly,
$$W=\frac{1}{2\pi}\int_{S^2}da. $$
The remarkable result in quantum field theory tells us that,
at nonzero values of $W$, the classical chiral symmetry cannot be promoted to quantum level. In fact,
\begin{equation}
Z_F\to e^{i\phi W}Z_F,
\label{4.4}\end{equation}
under the transformation (\ref{4.3}). This manifests the presence of a quantum anomaly.
In view of the well known index theorem, the winding number $W$ is given by the difference of positive and negative
chirality
zero  modes of the Dirac equation,
\begin{equation}
W=n_{+}-n_{-}.
\label{4.5} \end{equation}
This implies that the Dirac operator
$\hat D=-\hat{\partial} \pm i\hat a$ has zero eigenvalues at $W\neq 0,$
which
kills the fermionic path integral in (\ref{4.2}). The theory survives only if $W=0$. In this case
$\theta$ then multiplies zero and nothing can depend on it.

The direct consequence of that observation is that the difference between integer and half integer spins goes away
and
the low-energy excitations in the spin sector of the (\ref{4.2}) model become massive.
This is in agreement with the mean-field theory of the $1D$ $t-J$ model obtained within
the slave-fermion framework \cite{arovas}.
However, in the exact excitation spectrum of the $t-J$ model,
both spin and charge excitations are gapless in $1D$, at any hole concentration
\cite{coll, ogata}.
Specifically, in the limit $J\ll t$ the  $1D$ Hubbard model reduces to a squeezed Heisenberg chain with an enlarged lattice
constant
$\tilde a=a/(1-\delta)$ and a renormalized superexchange coupling $$\tilde J=J(1-\delta)\left(1-\frac{\sin 2\pi(1-\delta)}
{2\pi (1-\delta)}\right).$$
For small momentum the energy varies linearly with momentum,  $\epsilon(k)\propto \tilde Jk, \, k\to 0,$ leading to a
linear-temperature
contribution to the low-temperature specific heat.
The predicted linear term in the specific heat has been experimentally observed in a $1D$
organic molecular solid \cite{epstein}. This system can be described in terms of the $1D$
Hubbard model with a transfer integral of $2.1\times 10^ {-2}$ eV and an effective Coulomb interaction
of $0.17$ eV.
The action (\ref{4.2}) predicts instead an exponential fall off of the specific heat
and hence does not capture the low-energy physics of $1D$ strongly correlated electrons.

On the other hand, within the representation (\ref{2.10})
Eq.(\ref{4.01}) is replaced with
\begin{equation}
Z^{t-J}_{\vec S_i=0}=\int \prod_{\vec p,\tau}D\bar\xi_{\vec p}(\tau)D\xi_{\vec p}(\tau)
\exp{\left(-\sum_{\vec p}\bar\xi_{\vec p}(\tau)\xi_{\vec p}(\tau)
-\sum_{\vec p}\int_0^{\beta}\bar\xi_{\vec p}(\tau)
(\partial_{\tau}-\epsilon_{\vec p})\xi_{\vec p}(\tau)d\tau\right)},
\label{4.6}
\end{equation}
where the above discussed shift of the arguments of the path-integral variables is implicitly understood.
The time-lattice computation of this integral reduces to Eq.(\ref{1.24}) with the parameter
$\overline\mu$ replaced by $\mu$. As a result,
the partition function (\ref{4.6}) is found to diverge. This divergency is of course readily compensated
when the spin dynamics is turned on. However, the fermionic degrees of freedom
taken alone cannot be considered as well-defined physical entities independently from spin variables.
They act as auxiliary degrees of freedom to
describe a composite object - a constrained electron.
This also means that we cannot use the standard theory to describe the
low-lying fermionic excitation as given by Eqs.(\ref{4.01}),(\ref{4.1}).
The arguments that lead to Eq.(\ref{4.4}) are no longer applicable. Accordingly,
the gapless low-lying spin excitations cannot be ruled out in the present case.

We have not yet discovered how to perform the path integral (\ref{2.10}) in the continuum limit
in an analytically trustworthily way.
Integration over the $\xi$ or $z$ variables separately from each other does not in general result
in a physically relevant effective action.
For example, the action (\ref{1.20})
corresponds to the constrained on-site Hamiltonian (\ref{single}).
Integrating out the fermionic degrees of freedom
results in the divergent expression given by Eq.(\ref{1.24}).
A strongly correlated electron system at finite doping
can hardly be represented solely in terms of the $su(2)$ spin operators.
This agrees with the observation that the
spin-spin correlators in the $1D$
$t-J$ model involve the conventional $su(2)$ spin operators necessarily
modified by fermionic "string" operators.
Weng et al. \cite{weng} have shown that the effects due to the squeezing and rearrangement of spin configurations
in $1D$ $t-J$ model can be included by  introducing a nonlocal fermionic "string" field.
These processes of first squeezing the
$t-J$ chain and then rearranging the spin configuration cannot be described perturbatively
and must be taken into account by introducing string-like fields.
Formally, this can be thought of as a manifestation of the fact that the constrained electron (Hubbard)
operator cannot be
split into a convolution of conventional fermion and $su(2)$ spin operators. The bare charge and spin
degrees of freedom are not independent
and interact with each other very strongly
due to the NDO constraint. Within the path-integral
approach, this statement reasserts itself in the appearance of the extra factor in the measure
of the path integral (\ref{2.10}).

\section{Conclusion}

Let us now summarize the main results of our work. If one assumes on some phenomenological
grounds that the low-energy
physics of a strongly correlated electron system allows for a description in terms of the conventional
fermions coupled to the conventional lattice spins, one quite naturally ends up with the action
first suggested by Shankar. Strong correlations are encoded there as follows.
The theory comprises the spinless fermions which, even in the absence of the direct spin-fermion
interaction, are coupled to the lattice spins through the Berry's phase potential $a_0$ as exhibited in
Eq.(\ref{2.122}). That interaction brings in a correlation between the fermionic and spin degrees of freedom due
to the NDO constraint. However, this doesn't seem to be enough to account for strong correlations to full extent.
The example given by
Eq.(\ref{e2}) tells us that this approach is not truly self-consistent.

Our derivation of the appropriate low-energy action directly from the microscopical $t-J$
model brings out an important deviation from the Shankar's theory. Although the corresponding actions
look formally identical,
the measures
in the path integral representations for the corresponding partition functions
as well as the nature of the germane fields are essentially
different. The path-integral representation of the $t-J$ partition function (\ref{2.10})
involves the measure explicitly modified due to the NDO constraint. That modification ensures that we work
in the physical reduced Hilbert space.
As a result, the fermionic and bosonic amplitudes in the action (\ref{2.9}) no longer
correspond to the true fermion and lattice spin degrees of freedom, but rather represent
a unique composed object that describes the constraint electron as a whole.
Consequently, the doped $t-J$ model does not admit a representation
in terms of spinless fermions coupled to the local AF fluctuations via
a compact $U(1)$ gauge field even in the low-energy limit.
The explicit resolution of the NDO constraint results in
a rather involved path-integral representation for the $t-J$ partition function (\ref{2.10}).
It is not yet evident how one can proceed with its direct calculation, except in some trivial cases.
This appears as an evident consequence of the strong coupling nature of the physical
low-energy excitations of a system of the constrained electrons.

Within the slave-particle approach, this problem
is reflected  in the lack of a controlled treatment for the
emergent gauge field that strongly couples holons to spinons.
The NDO constraint which actually gives rise to that
gauge field \cite{lee2} is of a crucial importance right in the
underdoped $(\delta \ll 1)$ region which, by no accident, is the most interesting region of the phase diagram.
It therefore brings in
intractable strong interactions in this region, which in turn makes analytical calculations
completely uncontrolled. Essentially, the gauge theory based on the slave-particle representation
examines fluctuations around a mean-field solution that corresponds to a mean-field treatment
of the NDO constraint.
In general, such a theory is not stable against quantum fluctuations and the self-energy
corrections are infinite due to the low-frequency gauge field fluctuations.
The slave particles are not truly present in the physical spectrum, and cannot be treated as
quasiparticles weakly coupled to the gauge field \cite{nayak}.

On the other hand,
the Hartree-Fock approximations in the fermionic path integral for the Hubbard model,
\begin{equation}
H_{tU}=-t\sum_{ij,\sigma}c^{\dagger}_{i\sigma}c_{j\sigma}+U\sum_in_{i\uparrow}n_{i\downarrow},
\label{5.1}\end{equation}
are known to
recover  successfully  some well-known mean-field approximations such as, e.g., Stoner magnetism \cite{nagaosa}.
Those approaches are essentially based on the Hubbard-Stratonovich decomposition
of the $U$-quartic term which preserves the global spin
rotation symmetry \cite{schulz}. The Hubbard-Stratonovich fields are
then considered within the mean-field theory, which basically amounts to a saddle-point approximation
accompanied  by an  integration over the
corresponding  Gaussian fluctuations around its saddle point solution. This is equivalent to a random-phase
approximation (RPA)
and it is normally controlled by an appropriately chosen $1/N_f$ expansion, where $N_f$ stands for the number of the relevant
field components. However, it is very unlikely that the hole dynamics in the strongly correlated regime, close to
half filling,
admits such a mean-field description. In fact, the physics of Nagaoka's phase can hardly be recovered within such
a mean-field theory \cite{boies}.

In our view, the most appropriate way to proceed is to address the problem
of the low-energy dynamics of the $t-J$ model directly in terms of the superfield
$(\xi,z)\in CP^{1|1}$ as dictated by Eqs. (\ref{1.110}-\ref{1.113}).
After all, it is this composite field that
represents the true physical degrees of freedom - the constraint electron excitations.
In this way, one is supposed to
end up with a sigma model with the $CP^{1|1}$ target space. In fact, the $2D$ nonlinear sigma models
with the $CP^{n+m-1|n}$ target superspaces have been discussed to
describe percolations, polymers as well as some other problems in statistical mechanics \cite{saleur}.
This program can be explicitly carried out for a supersymmetric $t-J$ model
that exhibits a global $SU(2|1)$ invariance
at $J=2t$. Given a ground state of the lattice $t-J$ model,
the low-energy dynamics of the fluctuations around that vacuum state
is supposed to be controlled by the nonlinear $CP^{1|1}$ sigma model.
However, in a physically relevant case, $t\gg J$,
the vacuum state is still unknown (in fact, there exist
different competing vacuum states depending on the doping regime they are associated with).
Therefore, it is still
not clear
what kind of low-energy action actually emerges in these cases.

To summarize, there is still no complete analytical
effective theory to describe
the low-energy dynamics for the underdoped Mott insulator in general.
Such a theory should account for a
simultaneous existence of a few
competing nontrivial features: short-range AF order, superconductivity,
uniform and modulated spin-liquid regime and a pseudogap phase \cite{pepin}. The
important ingredient behind that picture is the local NDO
constraint that essentially affects the low-energy physics
close to half filling, and no trustworthy mean-field treatment is yet available
to tackle this problem.

\section{acknowledgments}
This research was supported in part by the Brazilian Ministry of Science and Technology (MCT) and
the Conselho Nacional de Desenvolvimento Cientifico e Tecnologico (CNPq).

\end{document}